\documentclass[twocolumn,separability,Amsterdam, nobibnotes, aps, pra,showpacs,preprintnumbers,nofootinbib]{revtex4-1}
\usepackage{amssymb}
\usepackage{array}
\usepackage{graphicx}
\usepackage{xcolor}
\usepackage{mathtools}
\usepackage{tensor}
\usepackage{graphicx}
\usepackage{amsmath}
\usepackage{amssymb}
\usepackage{mathrsfs}
\usepackage{color}
\usepackage{cancel}
\usepackage{tikz}
\usepackage{amsmath}
\usepackage{leftidx}

\def\beq{\begin{equation}}
\def\eeq{\end{equation}}
\def\bea{\begin{eqnarray}}
\def\eea{\end{eqnarray}}

\begin{document}

\title{Electromagnetic self-force in the five dimensional Myers-Perry space time}%

\author{Hamideh Nadi}
\email{h.nadi@ph.iut.ac.ir}
\author{Behrouz Mirza}
\email{b.mirza@cc.iut.ac.ir}
\affiliation{Department of Physics,
Isfahan University of Technology, Isfahan 84156-83111, Iran}
\author{Zahra Mirzaiyan} 
\email{z.mirzaiyan@ph.iut.ac.ir}
\affiliation{Department of Physics,
Isfahan University of Technology, Isfahan 84156-83111, Iran}
\affiliation{Erwin Schr\"{o}dinger Institute for Mathematics and Physics (ESI), Boltzmanngasse 9A, 1090 Wien , Austria}

\begin{abstract}
We calculate the effects of the electromagnetic self-force on a charged particle outside a five dimensional Myers-Perry space-time. Based on our earlier work \cite{zahra}, we obtain the self-force using quaternions in Janis-Newman and Giampieri algorithms. In four dimensional rotating space-time the electromagnetic self-force is repulsive at any point, however, in five dimensional rotational space-time, we find a point $r_0$ where the electromagnetic self-force vanishes. For $r<r_0 $ ($r>r_0 $) the electromagnetic self-force is attractive (repulsive).
\end{abstract}


\maketitle

\section{Introduction}

The interesting subject of self-force was proposed for the first time in the seminal work by Dirac in which the motion of the electron in flat space-time was studied \cite{Dirac}. The Dirac's work was generalized by DeWitt and Brehme in 1960 for the motion of a point particle in curved space-time \cite{Dewitt} in three cases: a scalar charge, an electric charge and a point mass in a curved space-time \cite{poisson}.
The so called ``self-force" comes from the interaction of a point particle with its own field. Although this force is usually very small and negligible, it deviates the point mass from moving on the background's geodesics.\\

 The gravitational self-force for a point particle outside the Schwarzschild space-time was calculated in \cite{Barack}.  For the case of Schwarzschild and Kerr space-times in four dimensions, the electromagnetic self-force for a static charge was obtained in a closed form \cite{Smith,BL}. Also in five dimensional black hole with the Schwarzschild-Tangherlini metric, the electromagnetic self-force for a charged particle was computed in \cite{poisson2}. For some recent works on self-force see \cite{JSW2015,Taylor2015,Ce2015,Har2016,ax,rec,poisson2018,cas2018}.
Recently a new approach was presented to find the effects of self-force (electromagnetic) on a charged particle in Kerr space-time based on the Janis-Newman (JN) algorithm \cite{Broccoili}. This result is for an arbitrary polar angle $\theta$ which is a more general result than \cite{BL}.  This new method, which has already been defined for the conversion of static metrics to rotating ones, simplifies the self-force calculations and the corresponding force in Kerr space-time is recovered successfully in \cite{Broccoili}. For more on Janis-Newman algorithm and Giampieri simplification see \cite{NJ,KN,Erbin,Erbin2,Erbin3,Erbin4,Giampieri}.
\\
In this paper, we obtain the electromagnetic self-force acting on a static charged particle in the five dimensional Myers-Perry space-time \cite{Myersperry} by applying the JN algorithm on the self-force in static space-time derived in \cite{poisson2}. Our work is based on our recent proposal \cite{zahra} for finding the five dimensional Myers-Perry black hole from the static solution using quaternion's algebra. The novelty of the recent proposal was using the quaternions' algebra for the first time for calculating the metric of rotating black holes in dimensions higher than four, especially the five dimensional Myers-Perry black hole. Our approach for the self-force in five dimensions, leads to a closed form for the electromagnetic self-force in 5D rotating space-time. Our result for the electromagnetic self-force is novel and there is no other calculation of the self-force in 5D rotating space-time, however, it is a conjecture and should be confirmed by other methods.\\
\\

This paper is organized as follows: In Section II we describe the JN algorithm and Giampieri simplification and derive the Kerr solution from the Schwarzschild metric. We also rewrite the results obtained in reference \cite{Broccoili} for the electromagnetic self-force acting on the static charged particle outside the Kerr space-time. In Section III, which is the main part of our paper, we obtain the electromagnetic self-force for the static charged particle in five dimensional Myers-Perry black hole by using quaternions' algebra. Finally, Section IV is devoted to the conclusions.


\section{The self-force of a static charge in Kerr background using Janis-Newman algorithm and Giampieri simplification} \label{S2}

Janis-Newman (JN) approach is a method for deriving the
rotating solution from the static one in four dimensions based on the original work in 1965 \cite{NJ}. To use this algorithm, one has to introduce a set of null tetrads which is usually a complicated process. Giampieri simplified the JN approach in 1990 \cite{Giampieri}. Using Giampieri's suggestion, null tetrads could be avoided and one can work with the background metric. More details may be found in \cite{Erbin2}. Starting from the Schwarzschild metric:\\

\begin{eqnarray}\label{usualsch}
ds^{2}=-f(r) dt^{2}+\frac{dr^2}{f(r)}+r^2d{\Omega}^{2}.
\end{eqnarray}

One can formulate the algorithm for deriving the Kerr metric in the following steps:

1. Transforming the metric to the Eddington-Finkelstein (EF) coordinates, where, $u=t-r_{*}$ and  $r_{*}=\int{dr (\frac{-g_{rr}}{g_{tt}})^{\frac{1}{2}}}$ ($du=dt- f(r)^{-1} dr$) with $f(r)=1-\frac{2M}{r}$. Schwarzschild metric in Eddington-Finkelstein coordinates  is as follows:

\begin{equation}\label{metricinu}
ds^{2}=-f(r) du^{2}-2 du dr+r^2d{\Omega}^{2}.
\end{equation}

2. Complexification of the coordinates $u=u^\prime+i a\ \cos \psi $ and $r=r^\prime-ia\ \cos \psi$. We obtain the following differentials:

\begin{eqnarray}\label{dudr}
du= du^{\prime}-ia\ \sin{\psi}\ d{\psi},\nonumber\\
dr= dr^{\prime}+ia\ \sin{\psi} \ d{\psi}.
\end{eqnarray}

Introducing the angle ${\psi}$, the  four dimensional space-time is embeded into a five dimensional complex space-time. \\


3. Complex transformation of $f(r)$. Since ${f}(r)$ depends on the coordinates $r$ and $\bar{r}$ (complex conjugate of $r$), due to the rule $( \frac{1}{r} \longrightarrow \frac{{\text{Re}} (r)}{|r|^2})$, ${f}(r)$ transforms as
 \begin{eqnarray}
{f}(r)=1-\frac{2M}{r} \longrightarrow \tilde{f}(r)=1-\frac{2M\ {\text{Re}} (r)}{|r|^2}. 
 \end{eqnarray}

4. Angle fixing. Using the ansatz $i d{\psi}=\sin{\psi}\ d\phi,\ \ {\psi}={\theta}$, helps to find a real metric as follows (by omitting the primes):

\begin{eqnarray}\label{kerr}
ds^{2}&=&(1-\frac{2M r}{\rho^2})(du-a\ \sin^{2} {\theta}\ d\phi)^{2}\nonumber\\
&&-2(du-a\ \sin^{2} {\theta}\ d\phi)(dr+a\ \sin^{2} {\theta}\ d\phi)\nonumber\\
&&+ {{\rho }^{2}}d{\Omega}^{2},
\end{eqnarray}
where we have defined
\begin{equation}\label{ro2}
{{\rho }^{2}}={{r}^{2}}+{{a}^{2}}{{\cos }^{2}}\theta.
\end{equation}

5. Go back to the Boyer-Lindquist coordinates by using transformations $du=dt^{\prime}-g(r) dr$ and $d{\phi}= d{\phi}^{\prime}-h(r) dr$. Applying the conditions ${{g}_{tr}}={{g}_{r{\phi }'}}=0$ functions $g(r)$ and $h(r)$ can be obtained as follows:

\begin{equation}\label{gh}
g=\frac{r^2 +a^2}{\Delta},\ \ h=\frac{a}{\Delta},\nonumber\\
\end{equation}

\noindent where ${\Delta}$ is introduced as
\begin{equation}
{\Delta} = \tilde{f}(r) {\rho}^2+a^2 \sin^2 {\theta}.
\end{equation}
Finally, we obtain the Kerr metric in the Boyer-Lindquist coordinates as follows (omitting the primes):
\begin{eqnarray}
d{{s}^{2}}&=&-\tilde{f}(r)d{{t}^{2}}+\frac{{{\rho }^{2}}}{\Delta }d{{r}^{2}}+{{\rho }^{2}}d{{\theta }^{2}}+\frac{{{\Sigma }^{2}}}{{{\rho }^{2}}}{{\sin }^{2}}\theta \ d{{\phi }^{2}}\nonumber\\
&&+2a(\tilde{f}-1)\ {{\sin }^{2}}\theta \ dtd\phi.
\end{eqnarray}

The electromagnetic self-force can also be derived using the above algorithm. Recently an original approach to compute the self-force  in Kerr space-time was presented in \cite{Broccoili}, where JN approach was used. Although there is not any proof
that the JN approach and Giampieri simplification can be used for deriving the electromagnetic self-force,
the corresponding force in Kerr space-time is recovered successfully. That means one can find the self-force in Kerr space-time from the self-force in Schwarzschild background, just by using JN algorithm on the forces. In the following part we review deriving the self-force in rotating background in four dimensions.\\

Considering  a static charged particle (with electric charge $e$) in the Schwarzschild background, the radial component is the only non-zero part of self-force and reads as follows \cite{poisson}

\begin{eqnarray}\label{selfsh}
\mathrm{f}^r=\frac{M e^2}{r^3} \left(1-\frac{2M}{r}\right)^{\frac{1}{2}}.
\end{eqnarray}

The self-force (\ref{selfsh}) can be written in Eddington-Finkelstein (EF) coordinates by using the tensorial transformation $\mathrm{f}_{EF}^\mu=\frac{\partial x^\mu}{\partial x^\nu} \mathrm{f}^\nu$ as 

\begin{eqnarray}
&&\mathrm{f}_{EF}^u=-\frac{M e^2}{r^3}  \left(1-\frac{2M}{r}\right)^{-\frac{1}{2}},\\
&&\mathrm{f}_{EF}^r=\frac{M e^2}{r^3} \left(1-\frac{2M}{r}\right)^{\frac{1}{2}}.
\end{eqnarray} 
Writing the one form $\mathrm{f}:=\mathrm{f}_{\mu}  dx^\mu$ in  Eddington-Finkelstein coordinates leads to (one can lower indices by using Eq. (\ref{metricinu}))

\begin{eqnarray}\label{finfin}
\mathrm{f}_{EF}=\frac{M e^2}{r^3}  \left(1-\frac{2M}{r}\right)^{-\frac{1}{2}} \ dr.
\end{eqnarray} 

At this stage one has to complexify the  Eddington-Finkelstein coordinates with some specified rules as 

\begin{eqnarray}\label{rule}
&&\frac{1}{r}\ \longrightarrow \ \frac{{\text{Re}}(r)}{|r|^2}=\frac{{\text{Re}} (r)}{\rho^2},\\
&&\frac{1}{r^3}\ \longrightarrow \ \frac{{\text{Re}} (r)}{|r|^4}=\frac{{\text{Re}} (r)}{\rho^4}.
\end{eqnarray}

\noindent where, $\rho^2$ is already introduced in Eq. (\ref{ro2}).\\
\\
Applying JN algorithm by using Eq.(\ref{dudr}), the self-force in Kerr space-time reads as follows

\begin{eqnarray}
\mathrm{f}=\frac{M e^2 r}{\rho^4}  \left(1-\frac{2M\ r}{\rho^2}\right)^{-\frac{1}{2}} \ (dr+a \ \sin^2 \theta \ d\phi).
\end{eqnarray}

The self-force in Boyer-Lindquist coordinates in Kerr background is obtained as

\begin{eqnarray}\label{selfforce4}
\mathrm{f}&=&\frac{M e^2r}{\rho^4}\left(1-\frac{2M\ r}{\rho^2}\right) ^{-\frac{1}{2}} \times\nonumber\\
&&  \left [\left(1-\frac{a^2 \ \sin^2 \theta}{\Delta}\right)dr+a \ \sin^2 \theta \ d\phi \right].
\end{eqnarray}
The absolute value of the self-force (\ref{selfforce4}) is defined as $\mathrm{\hat{f}}=\pm (\mathrm{f}_\mu
\mathrm{f}^\mu)^\frac{1}{2}$. The absolute value of the self-force is obtained as

\begin{eqnarray}\label{abs}
\mathrm{\hat{f}}=\frac{M e^2 r}{\rho^4}=\frac{M e^2 r}{(r^2 +a^2 \cos^2 \theta)^2},
\end{eqnarray}

\noindent where, for the case of a static particle located in $\theta=0$, the above relation reduces to the self-force calculated in \cite{BL}. Equation (\ref{abs}) is depicted in Fig.(1). As it can be seen from the diagram, there is a peak in the plot that appears in $r_{{\max}}=\frac{a \cos \theta}{\sqrt{3}}$.

 \begin{figure}\label{fig1}
\includegraphics[width=9cm]{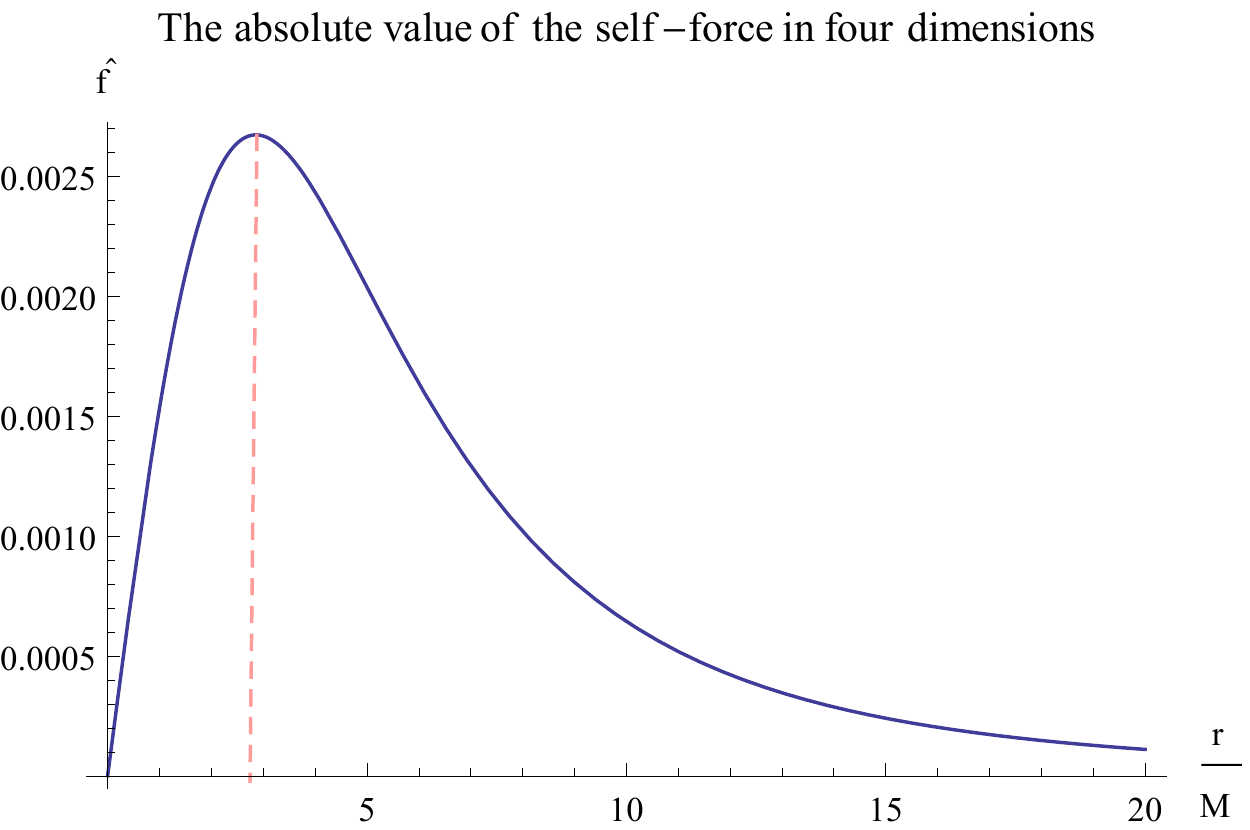}\\
 \caption[] {{Absolute value of self-force in four dimensions as a function of distance $r$ for $e=1$, $a=6$ and $\theta=0.6$. The peak in the plot shows where the maximum value for self-force for a static particle located at $r=2.8M$.}}\label{figure:add}
 \end{figure}

\section{The electromagnetic self-force of a static charged particle in five dimensional Myers-Perry space-time}\label{S4}

In the following, we review deriving the Myers-Perry black hole with two distinct angular momenta in a five dimensional space-time using quaternions \cite{zahra}. We derive the electromagnetic self-force of a static charged particle in 5-dimensional rotating space-time for the first time.\\
Starting with the 5-dimensional Schwarzschild solution as:
\begin{equation}\label{5sch}
ds^{2}=-f(r) dt^{2}+{f(r)}^{-1} dr^2+r^{2} d{\Omega}^{2} _{3},
\end{equation}

\noindent where, $f(r)=1-\frac{M}{r^{2}}$ and $d{\Omega}^{2} _{3}$ is the induced metric on the three dimensional sphere $S^{3}$ in the Hopf coordinates which can be written as follows:

\begin{equation}\label{domeg}
d{\Omega}^{2} _{3}=d{\theta}^{2}+ \sin^{2} {\theta}\ d{\phi}^{2}+ \cos^{2} {\theta}\ d{\psi}^{2},
\end{equation}

We repeat the same steps used in Sec. (\ref{S2}).\\

1. The Myers-Perry metric in five dimensions in Eddington-Finklestein retarded null coordinates can be written as:

\begin{equation}\label{newm}
ds^{2}=-du \  (du+2dr)+(1-f(r)) du^{2}+r^{2} d{\Omega}^{2} _{3}.
\end{equation}

2. Complexification of coordinates $u=u^\prime +ia \ \cos{{\xi}_{1}}+jb \ \sin{{\xi}_{2}}$ and $r=r^\prime -ia \ \cos{{\xi}_{1}}-jb \ \sin{{\xi}_{2}}$ using quaternions. Therefore, we obtain the following differentials 

\begin{eqnarray}\label{nj5}
du&=&du^{\prime}-ia\ \sin{{\xi}_{1}}\ d{{\xi}_{1}} +j b\ \cos{{\xi}_{2}}\ d{{\xi}_{2},}\nonumber\\
dr&=&dr^{\prime}+ia\ \sin{{\xi}_{1}}\ d{{\xi}_{1}} - j b\ \cos{{\xi}_{2}}\ d{{\xi}_{2}.}\nonumber\\
\end{eqnarray}

In (\ref{nj5}), $i$ and $j$ are the orthogonal basis of quaternions. Also, $a$ and $b$ are defined as parameters related to independent angular momenta. We may embed our five dimensional metric into a seven dimensional complex space-time.\\

3.  Complex transformation of $f(r)=1-\frac{M}{r^2}$. $f(r)$ depends on $r$ and its complex conjugate $\bar{r}$. Under the defined transformations, $f(r)$ transforms as (Note that $\frac{1}{r^2} \longrightarrow \frac{1}{|r|^2}$)

 \begin{eqnarray}\label{ftr}
&&{f}(r)=1-\frac{M}{r^2} \longrightarrow \tilde{f}(r)=1-\frac{M}{|r |^2}\nonumber\\
&&=1-\frac{M}{{r^\prime}^2 +a^2 \ \cos^2 \theta +b^2 \ \sin^2 \theta}. 
 \end{eqnarray}

4. Angle fixing.

\begin{eqnarray}\label{GA}
i d{{\xi}_{1}}&=&\sin{{\xi}_{1}} \ d{\phi},\ \ {{\xi}_{1}}={\theta},\nonumber\\
j d{{\xi}_{2}}&=&-\cos{{\xi}_{2}} \ d{\psi},\ \ {{\xi}_{2}}={\theta}.
\end{eqnarray}

Substituting (\ref{nj5}) and (\ref{ftr}) in metric (\ref{newm}) and by using angle fixing (\ref{GA}), we obtain the following form for the transformed metric

\begin{eqnarray}\label{trametric} 
ds^{2}&=&-(du^{\prime}-ia\ \sin{{\xi}_{1}}\ d{{\xi}_{1}} +j b\ \cos{{\xi}_{2}}\ d{{\xi}_{2}}) \times\nonumber\\
&& [(du^{\prime}-ia\ \sin{{\xi}_{1}}\ d{{\xi}_{1}} +j b \ \cos{{\xi}_{2}}\ d{{\xi}_{2}})+2(dr^{\prime}\nonumber\\
&&+ia\ \sin{{\xi}_{1}}\ d{{\xi}_{1}} - j b\ \cos{{\xi}_{2}}\ d{{\xi}_{2}})]\nonumber\\
&& +(1-{\tilde{f}}(r^{\prime})) (du^{\prime}-ia\ \sin{{\xi}_{1}} d{{\xi}_{1}} +j b \ \cos{{\xi}_{2}}\ d{{\xi}_{2}})^{2}\nonumber\\
&& +\mbox{angular part of the metric}.
\end{eqnarray}

Using the symmetrizing angle part method and the fact that quaternions are not commutative $(i \cdot j=-j \cdot i)$, we can now obtain the transformed metric as follows (all the primes are omitted). See Appendixes A and B for more details.
\begin{eqnarray}\label{lastmetric}
ds^{2}&=&-du^{2} -2 du dr \nonumber\\
&&+(1-\tilde{f} {(r)}) (du-a\ \sin^{2} {\theta}\ d{\phi} -b\ \cos^{2}{\theta}\ d{\psi} )^{2} \nonumber\\
&&+2a\ \sin^{2} {\theta}\ dr d{\phi}+2b\ \cos^{2} {\theta}\ dr d{\psi}+ {\rho}^{2} d{\theta}^{2}\nonumber\\
&&+(r^{2} +a^{2}) \sin^{2} {\theta}\ d{\phi}^{2}+(r^{2}+b^{2}) \cos^{2} {\theta}\ d{\psi}^{2}, \
\end{eqnarray}

\noindent where, $\rho^{2}=r \bar{r}=r^{{\prime}{2}} +a^{2} \cos^{2} {\theta}+b^{2} \sin^{2} {\theta}$.\\

5. For going to the Boyer-Lindquist coordinates, we use the following transformations:

\begin{eqnarray}\label{BLT}
du&=& dt-g(r) dr,\nonumber\\
d\phi &=& d{{\phi}^{\prime}}- h_{\phi}(r) dr,\nonumber\\
d\psi &=& d{{\psi}^{\prime}} -h_{\psi}(r) dr,\nonumber\\
\end{eqnarray}
where,
\begin{eqnarray}\label{hhg}
g(r)&=&\frac{\Pi}{\Delta},\nonumber\\
h_{\phi}(r)&=&\frac{\Pi}{\Delta} \frac{a}{r^{2}+a^{2}},\nonumber\\
h_{\psi}(r)&=&\frac{\Pi}{\Delta} \frac{b}{r^{2}+b^{2}},
\end{eqnarray}
Based on the definition of $\Pi =({r^{2}+a^{2}})({r^{2}+b^{2}})$ and $\Delta =r^{4} + r^{2} (a^{2} +b^{2}-m)+a^{2} b^{2}$,
we can obtain the five dimensional Myers-Perry solution in the Boyer-Lindquist coordinates as follows (omitting the primes):

\begin{eqnarray}\label{fmetric}
ds^{2}&=&-dt^{2} +(1-\tilde{f}(r))(dt-a\ \sin^{2} {\theta}\ d{\phi} -b\ \cos^{2}{\theta}\ d{\psi} )^{2}
\nonumber\\
&&+{\rho}^2 d{\theta}^{2} +\frac{r^{2} {\rho}^{2}}{\Delta} dr^2+(r^{2} +a^{2}) \sin^{2} {\theta}\ d{\phi}^{2}+ \nonumber\\
&&+(r^{2}+b^{2}) \cos^{2} {\theta}\ d{\psi}^{2}.
\end{eqnarray}

Since we know how to find the Myers-Perry metric from the non-rotating one in five dimensions with the proper algorithm,
we may obtain the electromagnetic self-force for a charged particle in five dimensional Myers-Perry space-time. Since the self-force in Kerr space-time can be derived successfully with this method \cite{Broccoili}, we expect our method leads to a correct form of self-force in five dimensional rotating background.\\

Considering a charged particle at the fixed position $r$ in the five dimensional Schwarzschild background (\ref{5sch}), only the radial component of self-force $F^r$, is non-zero and is given by \cite{poisson2}:

\begin{equation}
F^r=\frac{e^2 R^2}{2 r^5}\frac{\Gamma(x)}{f},
\end{equation}

\noindent where, $R$ is the radius of event horizon that is related to ADM mass $R=\sqrt{M}$, $f=1-(\frac{R}{r})^2$, $e$ is the electric charge and,

\begin{eqnarray}\label{Xi}
\Gamma(x) &=&-\frac{1}{4x}+\frac{5}{8}+\frac{139}{96} x -\frac{281}{192} x^2 +(\frac{1}{4x}+\frac{1}{2}-\frac{15}{16}x) \sqrt{f}\nonumber\\
&&+\frac{3}{16} x (6-5x) \ln(\frac{\bar{s} x (1+\sqrt{f})}{8 \sqrt{f}}),
\end{eqnarray}

Also $x=(\frac{R}{r})^2$ and $\bar{s}=\frac{s}{R}$. The self-force is dependent on an unidentified parameter $s$ which it can be interpreted as the radius of the charged particle.\\

We can obtain two non-zero components for the self force in Eddington-Finkelstein (EF) coordinates as

\begin{eqnarray}
&&F_{EF}^r=\frac{e^2 R^2}{2 r^5}\frac{\Gamma(x)}{f},\\
&&F_{EF}^u=-\frac{e^2 R^2}{2 r^5}\frac{\Gamma(x)}{f^2}.
\end{eqnarray}

The one form $\mathrm{F}=F_{\mu}dx^\mu$ in Eddington-Finkelstein coordinates is written as (one can lower indices by using Eq. (\ref{newm}))

\begin{eqnarray}\label{form}
\mathrm{F}_{EF}=\frac{e^2 R^2}{2 r^5}\frac{\Gamma(x)}{f^2} dr.
\end{eqnarray}

To use the Janis-Newman algorithm, we need the following specified rules

\begin{equation}
\frac{1}{r^2} \ \longrightarrow \ \frac{1}{|r|^2}=\frac{1}{\rho^2},\\ \label{com}
\end{equation}
\begin{equation}\label{com2}
\frac{1}{r^5} \ \longrightarrow \ \frac{\text{Re} (r)}{|r|^6}=\frac{\text{Re} (r)}{\rho^6},
\end{equation}

\noindent where, $\rho^2 = r^2 +a^2 \ \cos^2 \theta+b^2 \ \sin^2 \theta$.\\

Applying the above rules along with Eqs. (\ref{nj5}) with the angle fixing ansatz (\ref{GA}), the self-force (\ref{form}) acting on a static charge in five dimensional Myers-Perry space-time reads as follows

\begin{eqnarray}\label{Xirot}
\mathrm{F} &=&(\frac{e^2 R^2  r }{2 \rho^6 \ {\tilde{f}}^2})\  \Gamma (\rho) \times\nonumber\\
&& (dr+ a \ \sin^2 \theta \ d \phi+ b \ \cos^2 \theta \ d\psi),
\end{eqnarray}

where, $\tilde{f}=1-\frac{R^2}{\rho^2}$ and 

\begin{eqnarray}
 \Gamma(\rho)&=&-\frac{\rho^2}{4 R^2}+\frac{5}{8}+\frac{139}{96} \frac{R^2}{\rho^2} -\frac{281}{192} \frac{R^4}{\rho^4}\nonumber\\
 && +(\frac{\rho^2}{4 R^2}+\frac{1}{2}-\frac{15}{16}\frac{R^2}{\rho^2}) \sqrt{\tilde{f}}\nonumber\\
&&+\frac{3}{16} \frac{R^2}{\rho^2} (6-\frac{5R^2}{\rho^2}) \ln\frac{\bar{s}\ \frac{R^2}{\rho^2} (1+\sqrt{\tilde{f}})}{8 \sqrt{\tilde{f}}}.
\end{eqnarray}

We can easily find the self-force in Boyer-Lindquist coordinates in five dimensional Myers-Perry space-time using Eqs. (\ref{BLT}) and (\ref{hhg}) as

\begin{eqnarray}\label{Xirot2}
\mathrm{F} &=&(\frac{e^2 R^2  r }{2 \rho^6 \ {\tilde{f}}^2}) \ \Gamma(\rho) \times\nonumber\\
&&[(1-\frac{\Pi}{\Delta} (\frac{a^2 \ \sin^2 \theta}{r^2 +a^2}+\frac{b^2 \ \cos^2 \theta}{r^2 +b^2}))dr\nonumber\\
&&+ a \ \sin^2 \theta \ d \phi+ b \ \cos^2 \theta \ d\psi ].
\end{eqnarray}
Through raising indices $F_{\mu}$ by using the metric (\ref{fmetric}), we can calculate absolute value of the self-force, $\hat{F}=\pm(F_{\mu} F^{\mu})^\frac{1}{2}$, as follows (Note that the sign of $\hat{F}$ is chosen such that it agrees with the sign of $\mathrm{F}^r$)

\begin{eqnarray}\label{abs5}
\hat{F}=\frac{e^2 R^2 r}{2 \rho^6} \frac{\Gamma(\rho)}{{\tilde{f}}^{3/2}},
\end{eqnarray}

The absolute value of self-force in five dimensions, Eq. (\ref{abs5}) is plotted in Fig.(2) as a function of distance. An interesting feature in five dimensions for the self-force is that for some specific values of angular momentum, there is a point ($r_0$) where the value of self-force vanishes. The self-force becomes attractive (repulsive) for $r<r_0$ ($r>r_0$), however in four dimensions the self-force is positive for all values of $r$. Therefore if we live in a five dimensional space-time the extra dimension may be explored using experimental methods\footnote{We would like to thank anonymous referee for suggesting this idea.}. Also for the specific case of $a=b=0$, Equation (\ref{abs5}) is consistent with the self-force calculated for the non-rotating black hole in \cite{poisson2}.\\

Moreover note that there is no explicit calculation for the absolute value of self-force of a charged particle in five dimensional rotating Myers-Perry space-time. Our proposal is based on JN algorithm and should be confirmed by other methods.

 \begin{figure}\label{fig2}
\includegraphics[width=9cm]{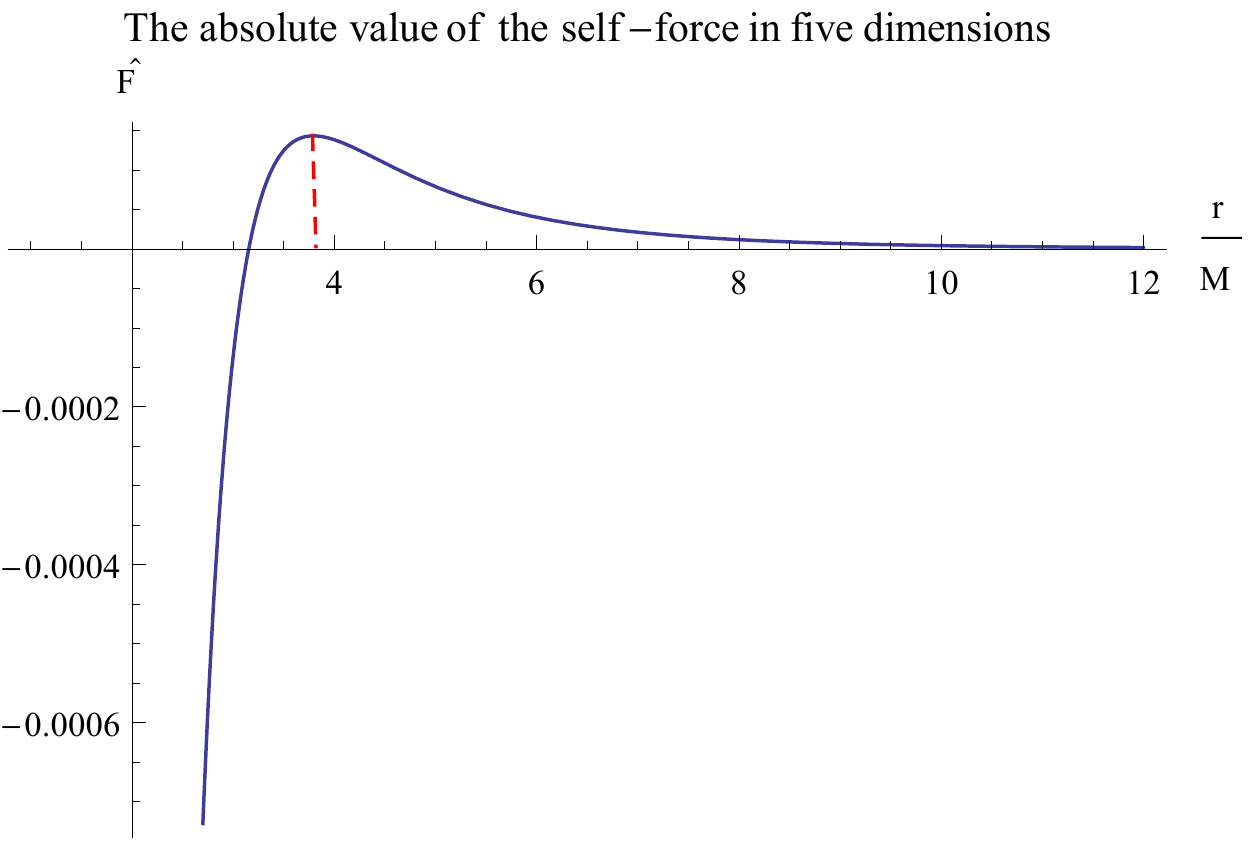}\\
 \caption[] {{Absolute value of self-force in five dimensions as a function of distance $r$ for $e=1$, $a=1$, $b=1$, $s=0.001$ and $\theta=0.6$. The peak in the plot shows where the maximum value for self-force for a static particle located at $r=3.8M$ and the self-force vanishes at $r=3.1M$. }}\label{figure:add}
 \end{figure}

\section{Conclusion}\label{S6}

In this work, we derived the self-force of a static charged particle in five dimensional Myers-Perry space-time. In this process, we used quaternion's algebra and their non-commutative property. We also showed that for specific values of angular momenta, self-force vanishes and make the particle move on the circular geodesics while this does not happens in four dimensions. This behavior might be used in the future experiments to measure self-force and therefore discovering possible extra dimensions. It will be very interesting to investigate this behavior in other examples and higher dimensions.

\section*{Acknowledgements:}
 ZM acknowledge the Erwin Schr\"{o}dinger Institute for Mathematics and Physics (ESI) scientific atmosphere. ZM was partially supported by the Erwin Schr\"{o}dinger JRF fund. 

\appendix

\section{Quaternions}

Quaternions are a system of basis in mathematics that are represented in a specific form as follows:

\begin{equation}\label{quater}
a+b\ i +c\ j+d\ k,
\end{equation}
\noindent where  $i, j,$ and $k$ are quaternion's units and $a,b,c,$ and $d$ are real numbers. Quaternion multiplications are presented in Table 1.\\
\begin{table}
\caption{Quaternion multiplication}
\begin{center}
\begin{tabular}{ | m{2em} | m{2em}| m{2em}| m{2em}|m{2em}|}
\hline
${\times}$ & 1 & i & j & k \\
\hline
1 & 1 & i & j & k \\
\hline
i& i & -1 & k & -j \\

\hline
j & j & -k & -1 & i \\

\hline
k & k & j & -i & -1\\

\hline
\end{tabular}
\end{center}
\end{table}

\section{Transforming the angular part of the five dimensional metric using symmetrizing angular terms and non-commutative feature of quaternions}
We show how the angular part in metric (\ref{trametric}) transforms using the symmetrizing method introduced in \cite{zahra}. By the angular part we mean $r^2 d{\Omega}^{2} _{3}=r^2 \ (d{\theta}^{2}+ \sin^{2} {\theta}\ d{\phi}^{2}+ \cos^{2} {\theta}\ d{\psi}^{2})$.\\

\noindent As the angular part in the metric need to be transformed, we show how each term transforms separately as follows\\
\\
\noindent 1.The first term, $r^{2} d{\theta}^{2}$: 
\begin{equation}\label{ft}
r^{2} d{\theta}^{2}\longrightarrow (r d\theta) ({r d\theta)^*} = (r^{{\prime}{2}} +a^{2} \cos^{2} {\theta}+b^{2} \sin^{2} {\theta})d{\theta}^{2}.
\end{equation}

\noindent 2. The second term, $r^{2} \sin^{2} {\theta} d{\phi}^{2}$:\\

\begin{eqnarray}\label{st}
r^{2} \sin^{2} {\theta} d{\phi}^{2}\longrightarrow \sin^{2} {\theta} (r d{\phi}) (r d{\phi})^{\ast}.
\end{eqnarray}

We replace the term $r d {\phi}$ by the following ansatz (see Eq.(\ref{GA})):
\begin{eqnarray}\label{dphi}
&&r d{\phi}= r (\frac{i d{{\xi}_{1}}}{\sin{{\xi}_{1}}})=(\frac{i\cdot r+r\cdot i}{2})( \frac{ d{{\xi}_{1}}}{\sin{{\xi}_{1}}}) \nonumber\\
&&=[\frac{1}{2} (i\cdot (r^{\prime}-i a\ \cos{{\xi}_{1}}-j b\ \sin {{\xi}_{2}})\nonumber\\
&&+(r^{\prime}-i a\ \cos{{\xi}_{1}}-j b\ \sin {{\xi}_{2}})\cdot i) \frac{d{{\xi}_{1}}}{\sin{{\xi}_{1}}}]\nonumber\\
&&=\frac{i d{{\xi}_{1}}}{\sin{{\xi}_{1}}} ( r^{\prime} -i a\ \cos{{\xi}_{1}}),
\end{eqnarray}

\noindent where, we used a symmetric form for quaternion's products. It should be noted that $(i \cdot j=-j \cdot i)$.\\

Using (\ref{dphi}), we can write (\ref{st}) as follows:
\begin{eqnarray}\label{st2}
&& r^{2} \sin^{2} {\theta} d{\phi}^{2}\longrightarrow  \ \ \sin^{2} {\theta}\ \ ( r^{\prime} -i a\ \cos{{\xi}_{1}}) ( r^{\prime} + ia\ \cos{{\xi}_{1}}) \frac{ d{{\xi}_{1}^{2} }}{\sin^{2} {{\xi}_{1}}}\nonumber\\
&& =\sin^{2} {\theta} (r^{\prime} -i a\ \cos{{\xi}_{1}}) (r^{\prime} + i a\ \cos{{\xi}_{1}}) \ d{\phi}^{2}\nonumber\\
&&= \sin^{2} {\theta} (r^{{\prime}{2}} + a^2 \cos^{2} {\theta})\ d{{\phi}^{2}},
\end{eqnarray}
where, $d{\phi}^2=\frac{ d{{\xi}_{1}^{2} }}{\sin^{2} {{\xi}_{1}}}$ is used in the third line and the substitution of  ${\xi}_{1}=\theta$ which is the angle fixing condition in (\ref{GA}).\\

\noindent 3. The third term, $r^{2} \cos^{2} {\theta} d{\psi}^{2}$:\\

\begin{eqnarray}\label{tt}
r^{2} \cos^{2} {\theta}\ d{\psi}^{2}\longrightarrow \cos^{2} {\theta}\ (r d{\psi}) (r d{\psi})^{\ast}.
\end{eqnarray}
Using the angle fixing, $d{\psi}=\frac{-j d{{\xi}_{2}}}{\cos{{\xi}_{2}}}$, we have

\begin{eqnarray}\label{tt2}
r^{2} \cos^{2} {\theta}\ d{\psi}^{2}\longrightarrow \cos^{2} {\theta}\ (r^{{\prime}{2}} +b^{2} \sin^{2} {\theta})\ d{{\psi}^{2}}.
\end{eqnarray}

Substituting Eqs. (\ref{ft}), (\ref{st2}) and (\ref{tt2}) in (\ref{trametric}) one can find the transformed metric (\ref{lastmetric}).




\end{document}